\documentclass[twocolumn,showpacs,preprintnumbers,amsmath,amssymb,prb,superscriptaddress]{revtex4-1}
\usepackage{mathrsfs}
\usepackage{graphicx}
\usepackage{dcolumn}
\usepackage{bm}
\usepackage{amsmath}
\usepackage{amsfonts}

\begin{document}

\title{Elastic scattering of surface states on three-dimensional topological insulators}
\author{Jing Wang}
\affiliation{State Key Laboratory of Low-Dimensional Quantum Physics, and Department of Physics, Tsinghua University, Beijing 100084, China}
\affiliation{Department of Physics, McCullough Building, Stanford University, Stanford, California 94305-4045, USA}
\author{Bang-Fen Zhu}
\thanks{To whom correspondence should be addressed.\\ bfz@mail.tsinghua.edu.cn}
\affiliation{State Key Laboratory of Low-Dimensional Quantum Physics, and Department of Physics, Tsinghua University, Beijing 100084, China}
\affiliation{Institute of Advanced Study, Tsinghua University, Beijing 100084, China}

\begin{abstract}
Topological insulators as new type of quantum matter materials are characterized by a full insulating gap in the bulk and gapless edge/surface states which are protected by time-reversal symmetry. We propose the interference patterns caused by elastic scattering of defects or impurities are dominated by surface states at the extremal points on the constant energy contour. Within such formalism, we summarize our recent theoretical investigations on elastic scattering of topological surface states by various imperfections, including the non-magnetic impurities, magnetic impurities, step-edges, and various other defects, in comparison with recent related experiments in typical topological materials such as BiSb alloys, Bi$_2$Te$_3$ and Bi$_2$Se$_3$ crystals.
\end{abstract}

\date{\today}


\maketitle

\section{Introduction}
\label{introduction}

The discovery of topological insulators (TIs) has attracted a great deal of attention.~\cite{hasan2010rmp,qi2011rmp}
A three-dimensional~(3D) TI has a time-reversal invariant band structure with nontrivial topological order, which possesses an energy gap in the bulk and gapless metallic states on the surface. The surface band consist of an \emph{odd} number of Dirac cones with unconventional spin texture in the surface Brillouin zone. One of the intriguing properties in such spin-textured surface states (SSs) is their insensitivity to spin-independent scattering, which is protected from backscattering by the time-reversal symmetry (TRS). Exotic excitations such as Majorana fermions,~\cite{fu2008prl} magnetic monopole~\cite{qi2009science} and dynamical axion field~\cite{li2010np} are predicted to exist as results of topological SSs.

Soon after the theoretical prediction~\cite{fu2007prb}, Bi$_{1-x}$Sb$_x$ alloy with composition $x$ ranged from $0.07$ to $0.22$ was found to be a TI with \emph{five} Dirac cones by the angle-resolved photoemission spectroscopy (ARPES).~\cite{hsieh2008nature}
ARPES measurements observed topologically nontrivial SSs, and the spin-resolved ARPES measurements reported the helical spin texture of the massless Dirac fermions in the system.~\cite{hsieh2009science} However, the SSs in Bi$_{1-x}$Sb$_x$ are rather complicated which can hardly be described by a simple model Hamiltonian. Subsequently, a family of binary compounds, Bi$_2$Se$_3$, Bi$_2$Te$_3$ and Sb$_2$Te$_3$, was discovered and investigated as the second generation of TIs.~\cite{xia2009np,zhanghj2009np,hsieh2009nature,chen2009science}  Among them the Bi$_2$Se$_3$ contains only a \emph{single} Dirac cone within the gap as large as $300$ meV, and as the ``hydrogen atom'' of TIs is easier to study. The helical spin texture of SSs in Bi$_2$Se$_3$ is observed to be \emph{left-handed}.~\cite{hsieh2009nature}
All these characteristics suggest that direct backscattering between the time-reversal pair of helicity states with opposite momenta and spins is prohibited, and the Anderson localization will not occur in these two-dimensional (2D) helical liquid.

In addition to the ARPES, the low-temperature scanning tunneling spectroscopy (STS) as well as microscope (STM) is another important probe in the TI research field, which provides a direct way to study the SSs through measuring the local density of states (LDOS) in the vicinity of various imperfections.
Due to the elastic scattering by impurities, the incoming surface wave with a wavevector
$\mathbf{k}^i$ must be scattered into the outgoing one with
$\mathbf{k}^f$ on the same constant energy contour (CEC). The quantum interference between the incoming and outgoing waves gives rise to the standing wave oscillation with wavevector $\mathbf{q}=\mathbf{k}^f-\mathbf{k}^i$,
 which is often referred as the characteristic scattering wavevector, or the transferred momentum.
Energy-resolved Fourier transform STS (FT-STS) can be used to obtain the scattering wavevectors of modulations in the LDOS and detailed information on the symmetry as well as
physics of scattering processes for the SS electrons.~\cite{petersen1998prb}  This technique has been proved valuable in studying the SSs on noble metals~\cite{crommie1993nature} and on semiconductor surfaces,~\cite{wittneven1998prl} and even on the pairing symmetry of the high-$T_c$ superconductors~\cite{hoffman2002science} and the graphene.~\cite{rutter2007science} The absence of backscattering is a specular manifestation of the topological nature of the SSs. Different from the edge states in two-dimensional (2D) TIs, in which the forbidden back-scattering means the annihilation of all types of elastic scattering by the spin-independent potentials, the SSs in 3D TIs will experience single-particle elastic scattering associated with all scattering angles except for $180^0$ backscattering.
Thus investigating the quasiparticle interference (QPI) caused by scattering off impurities and defects on the TI surfaces and revealing the topological nature of the SSs through the FT-STS are interesting and useful, in particular based on combination of the theoretical and experimental investigations.

The QPI patterns induced by non-magnetic impurities on surfaces of
Bi$_{x}$Sb$_{1-x}$,~\cite{roushan2009nature}
Bi$_2$Te$_3$~\cite{zhangtong2009prl} and Bi$_2$Se$_3$,~\cite{beidenkopf2011np} together with subsequent theoretical
analysis,~\cite{zhou2009prb,lee2009prb,wang2010prb,guo2010prb,biswas2011prb,wang2011prb}
did demonstrate the absence of backscattering for the topological SSs.
Meanwhile, the LDOS oscillations of SSs near step edges on
Bi$_2$Te$_3$ showed a power-law decay with
index $-3/2$, $-1/2$ and $-1$ for SSs at different energy ranges,~\cite{wang2011prb,alpichshev2010prl} while in Bi$_2$Se$_3$ the decay index is fixed to be $-3/2$,~\cite{wang2011prb,zhang2013prb}
compared to $-1/2$ for the two-dimensional electron
system~(2DES).~\cite{crommie1993nature} The faster decay of QPI pattern is believed to result from
the suppression of backscattering in TIs. For magnetic impurities on Bi$_2$Te$_3$ which breaks the TRS, although the TR violating scattering vectors have been reported,~\cite{okada2011prl}
it is still controversial whether the direct backscattering between the time-reversal pair is responsible for the unusual sensitivity to magnetic scattering.~\cite{beidenkopf2011np}

So far intensive STM and ARPES investigations have demonstrated the topological properties of SSs; however, a general picture of the QPI on the surface of various TI systems remains elusive, particularly when taking account of the quality of
TI samples and the warping effect of the Dirac cone.~\cite{chen2009science,fu2009prl} Here, not only the anisotropic band dispersion of the SSs, but also the warping effect includes coupling between the surface and bulk bands.
In this article we will give a comprehensive review on elastic scattering of SSs in various TI systems. The SS-LA-phonon quasi-elastic scattering plays an important role in transport of SSs,~\cite{huang2012} but not affect the QPI pattern significantly, so it will not be reviewed in this article.  With the recent experimental results, we present a general formalism to account for the complex QPI on TI surface.
We propose that the interference patterns are dominated by SSs at the extremal points on the CEC. In applying this theory to Bi$_2$Te$_3$ with strong warping effect, we show that, when tuning the bias voltage, for scattering off a non-magnetic impurity the QPI wavevector depends on the shape of Fermi surface sensitively, which varies from conical to $\bar{\Gamma}$-$\bar{K}$ and finally to $\bar{\Gamma}$-$\bar{M}$, as the CEC changing from circle to hexagon and finally to snowflake.
And the decay index off a step edge also critically depends on the energy of SSs,
varying from $-3/2$ to $-1/2$ and finally to $-1$ as the energy increases. As for TIs with nearly
isotropic Dirac cone, such as Bi$_2$Se$_3$,~\cite{kuroda2010prl} according to our theory the
decay index is simply $-3/2$. For the scattering off a magnetic impurity, the direct backscattering is allowed due to the TRS breaking. However, we will show such backscattering between time-reversal pairs in the presence of TRS breaking can be \emph{hardly} seen in LDOS by STM. Therefore, the QPI pattern of magnetic impurity is still similar to that of non-magnetic impurity.

The organization of this paper is as follows. After this
introductory section, Sec.~\ref{theory} describes the general theory for QPI of 2D SSs.
Here we focus on the Bi$_2$Se$_3$ family for their SSs can be described by a simple Hamiltonian.
Section~\ref{stepedge} is devoted to the standing wave due to step edge scattering compared to experiments.
Section~\ref{nonmagnetic} describes the QPI pattern by scattering off the non-magnetic impurity in combination with related experiments. Section~\ref{magnetic} describes the scattering of SSs by the magnetic impurity.
Section~\ref{otherimpurity} summarizes the experimental studies on the scattering off surface fluctuation and one-dimensional buckling. Section~\ref{conclusion} concludes this paper with a short summary and discussions.

\section{General Theory}
\label{theory}

\begin{figure}[t]
\begin{center}
\includegraphics[width=3.0in]{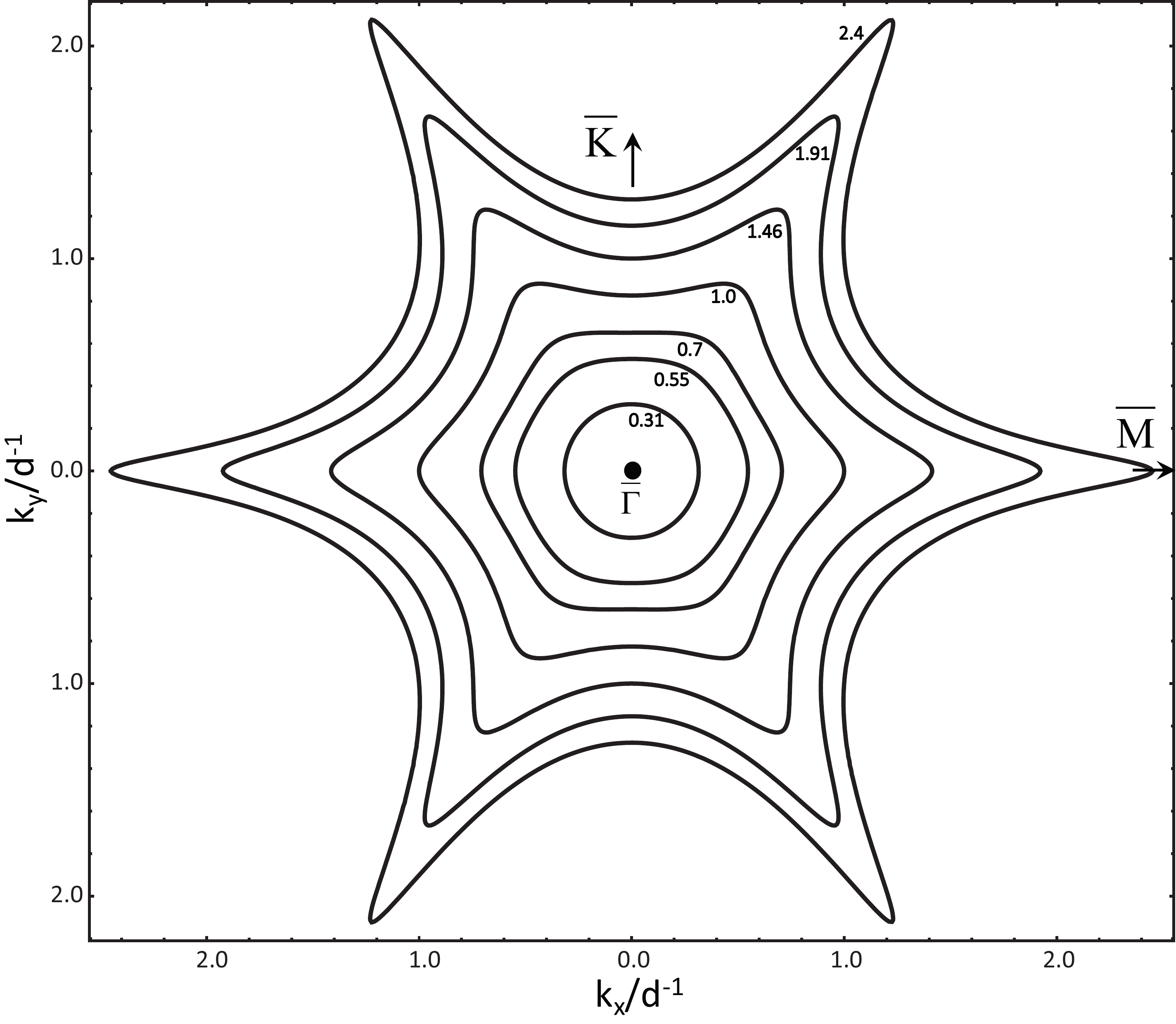}
\end{center}
\caption{Unified contours of constant energy and the evolution of Fermi surface of topological SSs in both Bi$_2$Se$_3$ and Bi$_2$Se$_3$, where $k_x$
and $k_y$ axis are in the unit of $\sqrt{v/\lambda}$, and energy is in unit of $\varepsilon^*$. The contours with BiSe parameters and $\varepsilon^*$ taken from $0.3$ to $0.7$ are suitable for Bi$_2$Se$_3$, and those with Bi$_2$Te$_3$ parameters and $\varepsilon^*$ from $0.7$ to $1.91$ are for BiTe}
\label{CEC}
\end{figure}

The topological SSs on Bi$_2$Te$_3$ and Bi$_2$Se$_3$
have a \emph{single} Dirac cone near $\bar{\Gamma}$ point in the surface Brillouin Zone on
each surface. The effective model describing such topological SSs
reads~\cite{fu2009prl}
\begin{equation}
\mathcal{H}_0(\mathbf{k}) = v(\sigma_xk_y-\sigma_yk_x)+\frac{\lambda}{2}\left(k_+^3+k_-^3\right)\sigma_z,
\end{equation}
where $\hbar\equiv1$, $k_{\pm}\equiv k_y\pm ik_x$, $v$ is the Dirac velocity, $\lambda$ is the warping parameter, and
$\sigma_i$ ($i=x,y,z$) are Pauli matrices acting on spin space. For simplicity,
we ignore the particle-hole asymmetry here as it affects the shape
of Fermi surface little. The surface band dispersion is
\begin{equation}
\varepsilon_{\pm}\left(k_x,k_y\right) = \pm\sqrt{v^2k^2+\lambda^2k^6\sin^2\left(3\theta\right)},
\end{equation}
where $\varepsilon_{\pm}$ denotes respectively the upper and the
lower energy bands touching at the Dirac point, and $\mathbf{k}
\equiv (k,\theta)$ with $\theta$ as the angle between the wave
vector $\mathbf{k}$ and $k_x$-axis~($\bar{\Gamma}$-$\bar{M}$). Defining the characteristic energy $\varepsilon^*\equiv
v\sqrt{v/\lambda}$ and length $d\equiv\sqrt{\lambda/v}$,
in Fig.~\ref{CEC} we plot a set of CEC of the upper band in momentum space
with \emph{no free parameters} for Bi$_2$Se$_3$ and Bi$_2$Te$_3$, respectively. For Bi$_2$Se$_3$,
$\lambda=128$~eV$\cdot${\AA}$^3$ and $\varepsilon^*=0.52$~eV,~\cite{xia2009np,kuroda2010prl} so that
the CEC is nearly isotropic from $0$ to
$0.42\varepsilon^*$~($0.22$~eV). As shown in Fig.~\ref{CEC}, when the Fermi
energy increases, the shape of CEC evolves from a
circle~($E=0.31\varepsilon^*$), more hexagon-like
($E=0.55\varepsilon^*$), hexagon ($E=0.7\varepsilon^*$) and to
concave hexagon ($E=1.0\varepsilon^*$). In Bi$_2$Te$_3$, $\lambda=250$~eV$\cdot${\AA}$^3$ and
$\varepsilon^*=0.23$~eV,~\cite{chen2009science} so the warping effect is stronger. As shown in
Fig.~\ref{CEC}, when the Fermi
energy increases, the shape of CEC evolves from a hexagon ($E=0.7\varepsilon^*$),
concave hexagon~($E=1.46\varepsilon^*$), more hexagram-like
($E=1.91\varepsilon^*$) and to six-pointed shape of snowflake ($E=2.4\varepsilon^*$).
The evolution of the Fermi surface with respect to energy agrees well with the ARPES results.~\cite{chen2009science,kuroda2010prl}
As we shall see, such changes in the shape of the Fermi surface will have drastic effects on the QPI around an impurity.

We consider the quasiparticle scattering problem
within the $T$-matrix approach. For a single non-magnetic or magnetic impurity located at the origin ($\mathbf{r}=0$) on the surface with the potential
\begin{equation}
\mathcal{H}_1 =V\delta(\mathbf{r})\sigma^{\mu},
\end{equation}
where $\sigma^{\mu}=1_{2\times2}$ for $\mu=0$ denotes non-magnetic impurity, and $\sigma^{\mu}=\sigma^{x,y,z}$ are the Pauli matrices
for $\mu=1,2,3$ denotes magnetic impurity. The LDOS can be expressed as
\begin{eqnarray}
\rho(\mathbf{r},\omega) &=& -\frac{1}{\pi}\mathrm{Im}\left[\mathrm{Tr}\left[G(\mathbf{r},\mathbf{r},\omega)\right]\right],\\
&=&\rho_0(\mathbf{r},\omega)+\delta\rho(\mathbf{r},\omega),\nonumber
\end{eqnarray}
where $\rho_0(\mathbf{r},\omega)$ is the LDOS of the unperturbed system with $V=0$, and $\delta\rho(\mathbf{r},\omega)$ is the deviation of the LDOS induced by the perturbation $H_1$.

Let the unperturbed surface Green's function be $G_0$, the perturbed Green's function can be expressed as
\begin{equation}
G(\mathbf{r},\mathbf{r}',\omega) = G_0(\mathbf{r}-\mathbf{r}',\omega)+G_0(\mathbf{r},\omega)T(\omega)G_0(-\mathbf{r}',\omega),
\end{equation}
with
\begin{eqnarray}
G_0(\mathbf{r},\omega) &=& \int^{\Lambda}\frac{d^2k}{(2\pi)^2}e^{i\mathbf{k}\cdot\mathbf{r}}G_0(\mathbf{k},\omega),\\
G_0(\mathbf{k},\omega) &=& \left[\omega+i\eta-\mathcal{H}_0(\mathbf{k})\right]^{-1},
\end{eqnarray}
where $\Lambda$ is the energy cutoff when integrating over momentum, and $T$ is given by
\begin{equation}
T(\omega) = V\sigma^{\mu}\left[1-G_0(\omega)V\sigma^{\mu}\right]^{-1}.
\end{equation}
Here we assume the translation symmetry in the unperturbed system and the momentum independent scattering potential. The spatial modulation of the LDOS is induced by $\mathcal{H}_1$ in the vicinity of impurities. To resolve the wave vector of the modulation of interference, it is convenient to calculate the Fourier-transformed LDOS
\begin{equation}
\delta\rho(\mathbf{q},\omega) = \int d^2r e^{-i\mathbf{q}\cdot\mathbf{r}}\delta\rho(\mathbf{r},\omega).
\end{equation}
From the real-space LDOS, one can get the power-law decay as a function of distance away from the impurity center $\delta\rho(\mathbf{r},\omega)\propto 1/r^{\alpha}$. While from the $q$-space LDOS, one can get the allowed wave vector $\mathbf{q}$ and the relative intensities for various scattering processes experienced by the SS electrons. Both the real-space and $q$-space information can be obtained by STM.

For the case of scattering by a step edge, the scattering potential is taken as $V\delta(x)$.
It is straightforward to compute the LDOS by the standard procedure described above.
However, without loss of generality, it is more convenient
to treat the scattering problem by using an analogy of
the elementary scattering problem with a barrier potential, either in
1D (step-edge) or 2D with rotational symmetry (magnetic or non-magnetic impurity), which is directly based on the wave function point of view.
We will discuss the formalism in Sec.~\ref{stepedgetheory} and Sec.~\ref{nonmagnetictheory}, respectively.

\section{Scattering by step-edge}
\label{stepedge}

\subsection{Theory}
\label{stepedgetheory}

Let us first consider the scattering of SSs by a step edge. The
step on a surface along a crystallographic axis in principle reduces
the interference problem to a 1D phenomenon. The step edge in a 2D conventional Fermi gas is known to give
rise to Friedel oscillation \emph{at a fixed energy} in the LDOS.~\cite{crommie1993nature}
Since the step edge scattering is a type of elastic scattering, the incoming SS with wave vector
$\mathbf{k}^i$ and the outgoing one with
$\mathbf{k}^f$ are on the same CEC. Assuming the step edge along the
$y$-direction, the $k_y$ component of the wave vectors should be
conserved in the scattering process, i.e. $k^i_y=k^f_y\equiv k_y$. The
interference between the incoming and outgoing waves  gives rise to
the standing wave oscillation in the $x$-direction. The total LDOS
is the sum of contributions from all these oscillations from the SSs
on a CEC.  For a given energy $E$, we can integrate over $k_y$ on
the entire CEC and express the LDOS explicitly as
\begin{equation}\label{integral}
\delta\rho(E,x) = \mathrm{Re}\left[\oint_E \frac{2r_c}{1+\left|r_c\right|^2}\xi_i^{\dag}\xi_fe^{i(k^f_x-k^i_x)x}dk_y\right],
\end{equation}
where $r_c$ is the reflection coefficient of the potential barrier of the step-edge, and $\xi$ denotes the spin
wave function of the SS of the form $\xi e^{ik_xx+ik_yy}$.
\begin{figure*}[tp]
\begin{center}
\includegraphics[width=5.0in]{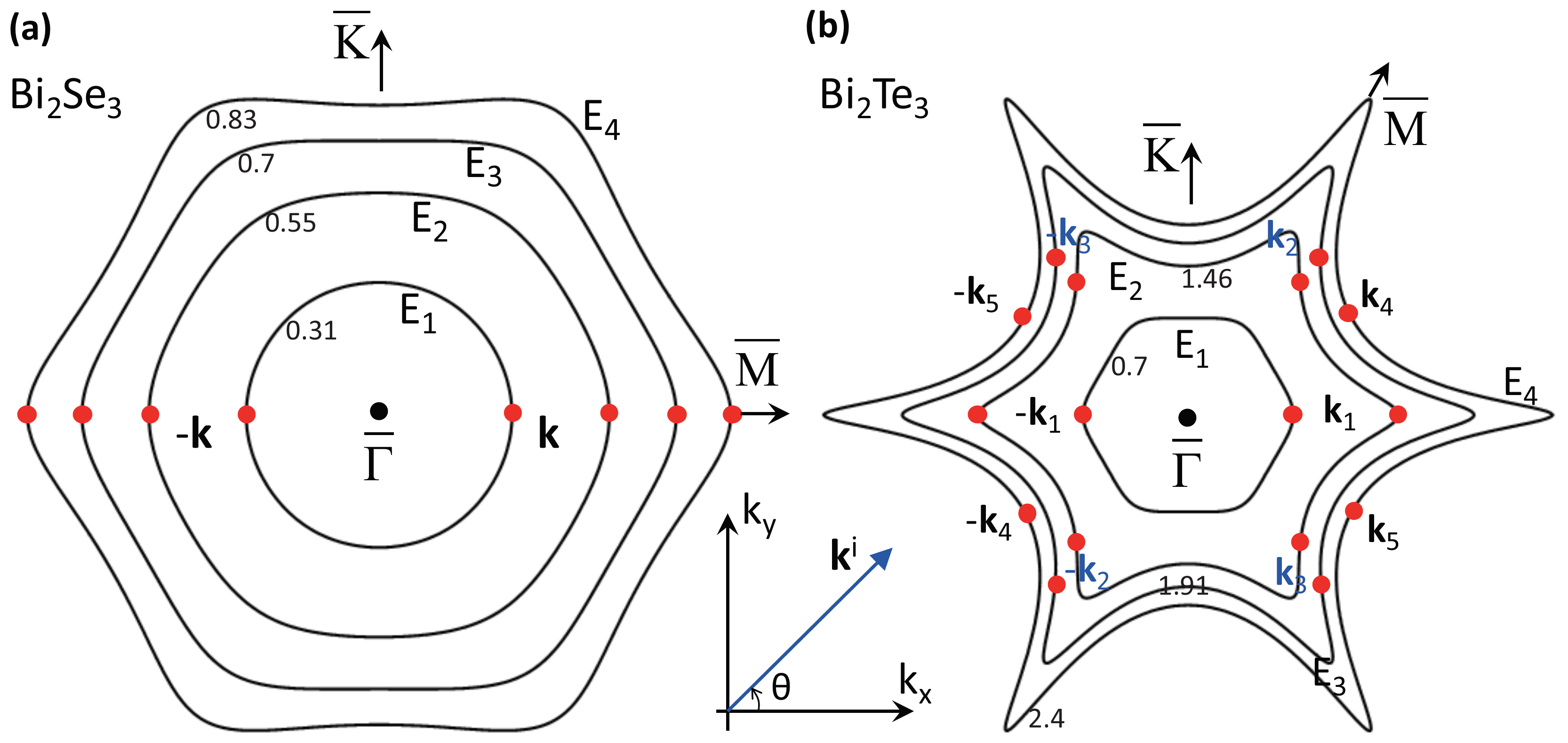}
\end{center}
\caption{(color online) Illustration of scattering off the step edge
along $y$-direction. The LDOS oscillation is dominated by scattering
between the EP pairs (red dots) on CEC. (a) The shape of CEC in
Bi$_2$Se$_3$ evolves from circle, hexagon to concave hexagon as
energy increases. It has a single pair of EPs at ($\mathbf{k}$,
$-\mathbf{k}$) and only one type of oscillation pattern appears with
different bias voltages. (b) In Bi$_2$Te$_3$, as energy increases,
CEC evolves from hexagon, concave hexagon to concave hexagram, corresponding to
the pair of EPs, first at ($\mathbf{k}_1$, $-\mathbf{k}_1$);
then at ($\mathbf{k}_1$, $-\mathbf{k}_1$), ($\mathbf{k}_2$,
$-\mathbf{k}_3$), ($\mathbf{k}_3$, $-\mathbf{k}_2$); then at
($\mathbf{k}_2$, $-\mathbf{k}_3$), ($\mathbf{k}_3$,
$-\mathbf{k}_2$); finally at ($\mathbf{k}_4$, $-\mathbf{k}_5$) and
($\mathbf{k}_5$, $-\mathbf{k}_4$). Different types of oscillation
pattern appear at different bias voltages.}
\label{edge}
\end{figure*}

A pair of states $\mathbf{k}^i$ and $\mathbf{k}^f$ scattered by the step along $y$-direction results in a standing wave with spatial period of
$2\pi/(k^f_x-k^i_x)\equiv 2\pi/\Delta k_x$. Since the oscillation period is
different for different value of $k_y$, a large number of scattering pairs will interfere destructively, and only the pair whose period
is \emph{stationary} with respect to small variation in $k_y$  may
make dominant contribution to the LDOS oscillations. Such
pair of points [($k^i_{x0}$,$k_{y0}$) and ($k^f_{x0}$,$k_{y0}$)] on
CEC are called the \emph{extremal points}~(EPs).~\cite{roth1955pr} Namely,
$\partial \Delta k_x /\partial {k_y}|_{k_{y0}}=0 $. And $\Delta k_{x0}$ is the characteristic scattering wavevector which critically depends on the geometry of CEC. Other standing
waves will cancel at large $x$ because of mutual interference. The
spatial dependence of LDOS oscillations in Eq.~(\ref{integral}) can
be evaluated by expanding the relevant quantities around each pair of EPs.
Let $k_y=k_{y0}+\delta k_y$, then $\Delta k_x=\Delta k_{x0} +
\sum_{n}\Delta k_{xn}\delta k_y^{n}$, $r=\sum_{l}\eta_l\delta
k_y^{l}$, and $\xi_i^{\dag}\xi_f=\sum_{m} \chi_m\delta k_y^{m}$.
To the leading order of $\delta k_y$, the LDOS varies
at long distance as
\begin{eqnarray}\label{LDOS}
\delta \rho(E,x) &\simeq& \mathrm{Re}\left[\sum\limits_{\mathrm{EPs}}\int_E2r/(1+\left|r\right|^2)\xi_i^{\dag}\xi_fe^{i(k_x^f-k_x^i)x}dk_y\right],
\nonumber
\\
&\sim& \sum\limits_{\mathrm{EPs}}\left|g\eta_{a}\chi_{b}c\right|\cos(\Delta k_{x0}x+\phi_s)x^{-\frac{a+b+1}{c}},
\end{eqnarray}
where $a$=$\min{(l)}$, $b$=$\min{(m)}$, $c$=$\min{(n)}$, $g=\oint_E
dk' k'^{(a+b-c+1)/c}e^{i\Delta k_{xc}k'}$, and $\phi_s$ is the
initial phase of each pair of EPs. The decay behavior of LDOS in
Eq.~(\ref{LDOS}) is valid as long as $x\gg\Delta k_{x0}^{-1}$. The
decay index associated with a pair of EPs is given by $(a+b+1)/c$,
which is solely determined by the properties of the scattering wave
function around the EPs on a given CEC.

The step edge is always along the close packed $\bar{\Gamma}$-$\bar{K}$
direction. As shown in Fig.~\ref{edge}(a), In Bi$_2$Se$_3$, when the Fermi
energy increases, the shape of CEC evolves from a
circle~($E_1=0.31\varepsilon^*$), more hexagon-like
($E_2=0.55\varepsilon^*$), hexagon ($E_3=0.7\varepsilon^*$) and to
concave hexagon ($E_4=0.83\varepsilon^*$). In a wide range of energy
only a single pair of EPs exists at $(\mathbf{k},-\mathbf{k})$, so
the characteristic wave vector is always equal to $2\mathbf{k}$ and
$c=2$. In Bi$_2$Te$_3$, the warping effect is stronger, and the power law decay behavior is more complicated. As shown in
Fig.~\ref{edge}(b), EPs evolve with the energy as follows: Single
pair of EPs ($\mathbf{k}_1$,$-\mathbf{k}_1$) at
$E_1=0.7\varepsilon^*$; Multiple pairs of EPs
($\mathbf{k}_1$,$-\mathbf{k}_1$), ($\mathbf{k}_2$,$-\mathbf{k}_3$)
and ($\mathbf{k}_3$,$-\mathbf{k}_2$) at
$E_2=1.46\varepsilon^*>E_c\equiv3^{1/3}\sqrt{11/9}\varepsilon^*\simeq1.45\varepsilon^*$;
Two pairs of EPs ($\mathbf{k}_2$,$-\mathbf{k}_3$) and
($\mathbf{k}_3$,$-\mathbf{k}_2$) survive at $E_3=1.91\varepsilon^*$,
as the SSs along the $\bar{\Gamma}$-$\bar{M}$ direction merge into
the bulk conduction band;  No EPs at all at $E_4=2.4\varepsilon^*$,
because the SSs separate from bulk states only in the very vicinity along
$\bar{\Gamma}$-$\bar{K}$ direction on the Fermi surface as observed
in the ARPES experiment.~\cite{chen2009science} In this case, 
 scattering between SSs around ($\mathbf{k}_4$,$-\mathbf{k}_5$)
and ($\mathbf{k}_5$,$-\mathbf{k}_4$) will mostly contribute to LDOS
oscillations, but they are not EPs and $c=1$. Thus in Bi$_2$Te$_3$ the characteristic wave vector and the LDOS oscillation period critically depend on the energy of SSs via varying the bias in the STS. In most
cases we have parameter $c=2$ except for Fermi energy as high as
$E_4$~($c=1$).

For the incoming state with wave vector
$\mathbf{k}^i$=$(k^i,\theta^i)$ and energy $\varepsilon_+(\mathbf{k}^i)$, the
inner product of two spin wave functions
$\xi_i^{\dag}\xi_f=\sin\theta^i+i\lambda(k^i)^3\sin(3\theta^i)\cos\theta^i/\varepsilon_+(\mathbf{k}^i)$.
It vanishes only for EPs associated with zero $\theta$, when the spins are exactly anti-parallel
for the time-reversal pairs of $(\mathbf{k},-\mathbf{k})$ and
$(\mathbf{k}_1,-\mathbf{k}_1)$. Since $b$ is the lowest power
for $\xi_i^{\dag}\xi_f$, for the time-reversal pairs $b\geq1$;
otherwise $b=0$. Thus, $b=1$ in Bi$_2$Se$_3$; while in
Bi$_2$Te$_3$, $b=1$ for the pair ($\mathbf{k}_1$,$-\mathbf{k}_1$),
and $b=0$ for other pairs of EPs.

Assuming the step edge potential
is $V(x)=0$ for $x<0$ and $V(x)=-V_0$ ($V_0>0$) for $x>0$,  by
matching the boundary condition at the edge the reflection
coefficient can be obtained as
\begin{equation}
r(\theta^i) = (-i)\frac{e^{-i(\theta^i-\theta^t)/2}-\beta(\theta^i)e^{i(\theta^i-\theta^t)/2}}
{e^{i(\theta^i+\theta^t)/2}+\beta(\theta^i)e^{-i(\theta^i+\theta^t)/2}},
\end{equation}
where $(k^t,\theta^t)$ is the momentum of the transmitted state,
\begin{equation}
\beta(\theta^i)=\frac{\varepsilon_+(\mathbf{k}^i)/k^i+\lambda(k^i)^2\sin(3\theta^i)}
{\varepsilon_+(\mathbf{k}^t)/k^t+\lambda(k^t)^2\sin(3\theta^t)},
\end{equation}
$\varepsilon_+(\mathbf{k}^i)=\varepsilon_+(\mathbf{k}^t)-V_0$, and
$\theta^t(\theta^i)=-\theta^t(-\theta^i)$. Due to the constraint by TRS, $r(\theta^i)=-r(-\theta^i)$, and the normal reflection $r(\theta^i=0)=0$,
which means the absence of backscattering. $a$ is the lowest power for $r$, 
thus $a=1$ for the backscattering pair of $(\mathbf{k},-\mathbf{k})$ in Bi$_2$Se$_3$ and
($\mathbf{k}_1$,$-\mathbf{k}_1$) in Bi$_2$Te$_3$, and $a=0$ for
other pairs in Bi$_2$Te$_3$.

In short, the algebraical decay index is $3/2$ for
$(\mathbf{k},-\mathbf{k})$ and ($\mathbf{k}_1$,$-\mathbf{k}_1$)
pairs, $1/2$ for ($\mathbf{k}_2$,$-\mathbf{k}_3$) and
($\mathbf{k}_3$,$-\mathbf{k}_2$) pairs, and $1$ for
($\mathbf{k}_4$,$-\mathbf{k}_5$) and
($\mathbf{k}_5$,$-\mathbf{k}_6$) pairs. Therefore, the LDOS
oscillations of the SSs in Bi$_2$Se$_3$ should decay as $x^{-3/2}$
in a wide range of energy (as long as $E<0.76$~eV), much faster than
$x^{-1/2}$ in 2DES.~\cite{crommie1993nature} On Bi$_2$Te$_3$
surfaces, as the bias increases, LDOS oscillations decay first as
$x^{-3/2}$ ($E<0.33$~eV), then as a combination of $x^{-3/2}$ and
$x^{-1/2}$, then as $x^{-1/2}$, and finally $x^{-1}$.

\begin{table}[htbp]
\begin{center}
\renewcommand{\arraystretch}{1.6}
\begin{tabular*}{6.0cm}
{@{\extracolsep{\fill}}cc}
\hline
\hline
\ &  decay power
\\ \hline
$0<E<1.45\varepsilon^*$  & $x^{-3/2}$ \\
$1.45\varepsilon^*<E<\varepsilon_1$  & $x^{-3/2}$, $x^{-1/2}$ \\
$\varepsilon_1<E<\varepsilon_2$  & $x^{-1/2}$ \\
$\varepsilon_2<E<\varepsilon_{\mathrm{max}}$  & $x^{-1}$ \\
\hline
\hline
\end{tabular*}\label{Tab2}
\end{center}\caption{Index of power law decay of standing waves of SSs scattering by step-edges. $\varepsilon_1$ denotes the energy at which the SSs along $\bar{\Gamma}$-$\bar{M}$ begin to merge into bulk conduction band, $\varepsilon_2$ denotes the energy where the SSs only exist in the very vicinity along $\bar{\Gamma}$-$\bar{K}$. $\varepsilon_{\mathrm{max}}$ is the energy upper bound for the surface Dirac electron.}
\end{table}

\subsection{Experiments}
\label{stepedgeexperiment}

Kapitulnik's group performed STM and STS study on Sn- and Cd-doped Bi$_2$Te$_3$ crystals.~\cite{alpichshev2010prl}
By analyzing the oscillation of LDOS near a step-edge,
they showed that topological SS oscillations are strongly damped compared to conventional SSs.
This is another hallmark of the severe suppression of backscattering,
hence supporting the TRS protected SSs in Bi$_2$Te$_3$. An interesting observation in experiments
is the emergence of pronounced oscillations with a distinct wave vector at higher energies, which may result from a
hexagonal warping of the surface band structure.

Xue's group reported the first interference fringes at the step edges on  Bi$_2$Se$_3$ surface thanks to their high quality sample.~\cite{zhangtong2009prl,wang2011prb}
By analyzing decay power of standing wave oscillations across a step on Bi$_2$Se$_3$, they showed that the oscillations decay index is $x^{-3/2}$, much faster than in 2D metal $x^{-1/2}$.~\cite{crommie1993nature} This confirms the suppression of backscattering of topological SSs in Bi$_2$Se$_3$ due to TRS. Such faster decay power of $-3/2$ in Bi$_2$Se$_3$ is also confirmed by Stroscio's group.~\cite{zhang2013prb}
While in Bi$_2$Te$_3$ with strongly warped surface band, they clearly reveal the variation of the QPI decay index with the bias voltage. As shown in Table 1, as the energy of SS increases, the power index varies from $x^{-3/2}$ to a mixture of $x^{-3/2}$ and $x^{-1/2}$, to $x^{-1/2}$, and finally to $x^{-1}$. The $x^{-1/2}$ decay indicates that the suppression of backscattering due
to TRS does not necessarily lead to a spatial decay rate faster than that in the normal 2DES. As shown in Ref.~\onlinecite{wang2011prb}, the agreement between experiment of Xue's group and our theoretical prediction is surprisingly perfect.

\section{Non-magnetic impurity}
\label{nonmagnetic}

\subsection{Theory}
\label{nonmagnetictheory}

\begin{figure*}[htbp]
\begin{center}
\includegraphics[width=5.0in]{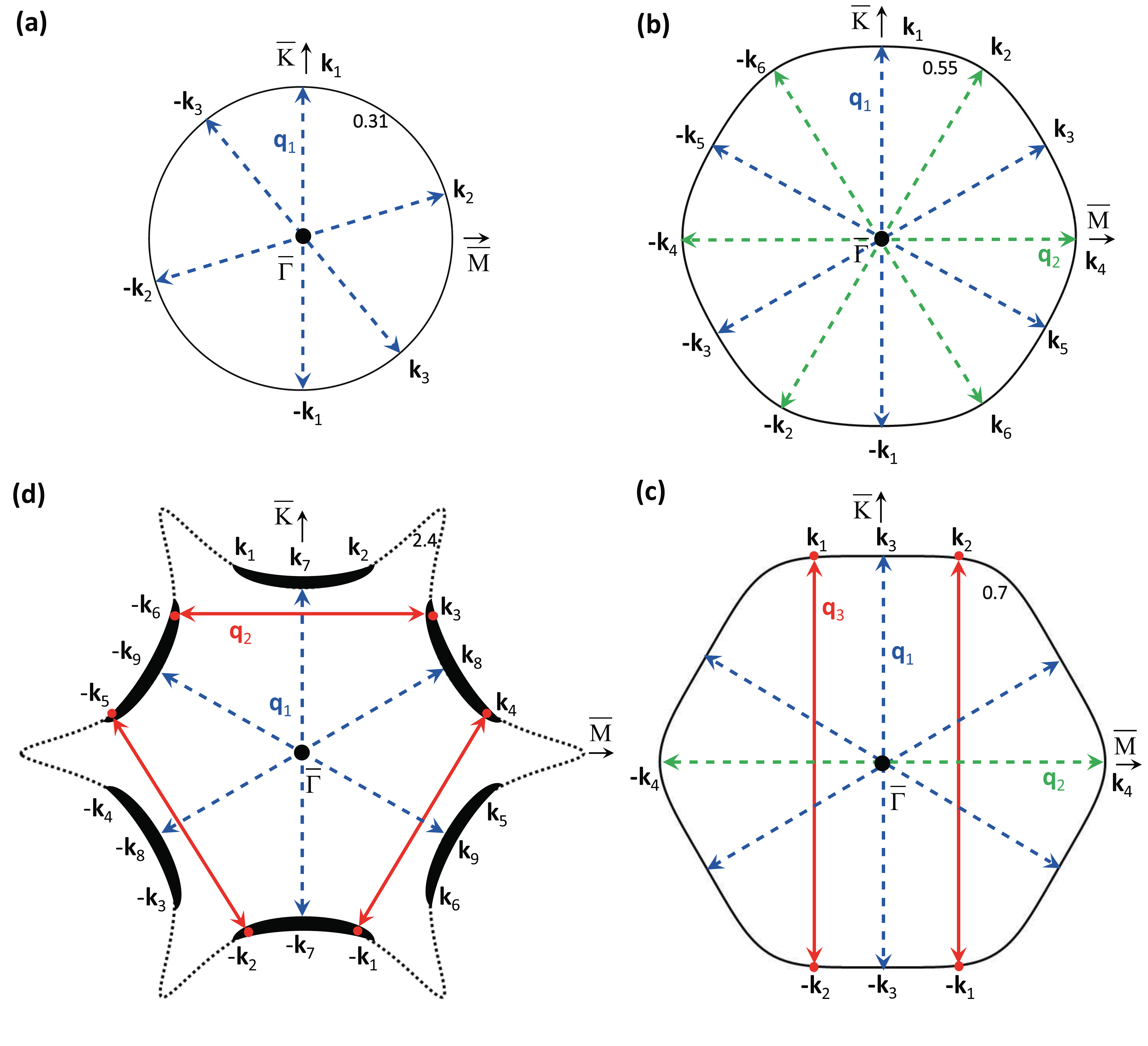}
\end{center}
\caption{(color online) Illustration of scattering geometry due to a non-magnetic impurity. The LDOS oscillation is dominated by scattering
between the pair of EPs on CEC. (a) For the CEC of circle,
the pair of EPs is the SS with $\mathbf{k}$ and its time-reversal state $-\mathbf{k}$, such as
($\mathbf{k}_1$, $-\mathbf{k}_1$), ($\mathbf{k}_2$, $-\mathbf{k}_2$) and ($\mathbf{k}_3$, $-\mathbf{k}_3$) etc. The QPI pattern in the $q$-space is circular.
(b) When the CEC evolves into convex, two types of forbidden characteristic wave vectors, $\vec{q}_1$ (dashed blue) and $\vec{q}_2$ (dashed green), connect the time-reversal pairs. (c) For the CEC of hexagon, $\vec{q}_1$ and $\vec{q}_2$ are forbidden, while $\vec{q}_3$ (solid red), in the direction of $\bar{\Gamma}$-$\bar{K}$, dominates the QPI pattern. (d) For the
hexagram CEC, the regions with high DOS are denoted by bold lines along the
direction of $\bar{\Gamma}$-$\bar{K}$. Two types of characteristic
scattering wave vectors include the forbidden $\vec{q}_1$ (dashed
blue) and the allowed $\vec{q}_2$ (solid red) which connects a
pair of EPs. Thus different oscillation
wave vectors appear in the FT-STS with different bias voltages.}
\label{impurity}
\end{figure*}

Now we switch from the 1D scattering to 2D scattering. We first consider the interference patterns of topological SSs due to a non-magnetic impurity. The interference wave vectors of the QPI pattern can be simply understood by exploring the standing wave formation mechanism. Suppose an incident wave $\psi_i\propto e^{i\mathbf{k}^i\cdot\mathbf{r}}/\sqrt{r}$ is scattered into
$\psi_f\propto f(\theta)e^{i\mathbf{k}^f\cdot\mathbf{r}}/\sqrt{r}$
by the impurity potential. The resultant LDOS oscillation is
\begin{equation}
\delta\rho(\mathbf{r},\omega)\propto \sum\limits_{\mathbf{k}^i,\mathbf{k}^f}\frac{1}{r}f(\theta)e^{i(\mathbf{k}^f-\mathbf{k}^i)\cdot\mathbf{r}},
\end{equation}
with $f(\theta)$ denoting the scattering amplitude and $r$ denoting the distance away from the impurity center. Also because of the elastic scattering, $\mathbf{k}^i$ and $\mathbf{k}^f$ are on the same
CEC. There are various scattering directions between incoming and outgoing SSs, but most of these processes interfere destructively, contributing little to the standing waves. The interference pattern are dominated by the pair of EPs on the CEC either,~\cite{wang2011prb} which resembles the Fermat's principle in optics. The characteristic
scattering wavevector of the QPI pattern at large distance in a given direction $\hat{r}$ comes from scattering
between states at EPs on the Fermi surface, where the Fermi velocity is parallel to $\hat{r}$.

For the circle-like CEC shown in Fig.~\ref{impurity}(a), the EP pair only exists at $\mathbf{k}$ and its time-reversal partner $-\mathbf{k}$. There are many such pairs, for example, ($\mathbf{k}_1, -\mathbf{k}_1$), ($\mathbf{k}_2, -\mathbf{k}_2$), and ($\mathbf{k}_3, -\mathbf{k}_3$), etc. Let $\vec{q}_1$ represent the backscattering vector between $\mathbf{k}$ and $-\mathbf{k}$. Obviously, there is no dominating direction of $\vec{q}_1$ in the circle case. The two states $\left|\mathbf{k},\uparrow\right\rangle$ and $\left|-\mathbf{k},\downarrow\right\rangle$, which are related by the TRS as $\left|-\mathbf{k},\downarrow\right\rangle=\mathcal{T}\left|\mathbf{k},\uparrow\right\rangle$, carry opposite spins, hence
\begin{eqnarray}\label{TRI}
\left\langle-\mathbf{k},\downarrow\right|\hat{U}\left|\mathbf{k},\uparrow\right\rangle &=& -\left\langle\mathbf{k},\uparrow\right|\hat{U}\left|-\mathbf{k},\downarrow\right\rangle^*
\nonumber
\\
&=&-\left\langle-\mathbf{k},\downarrow\right|\hat{U}\left|\mathbf{k},\uparrow\right\rangle = 0.
\end{eqnarray}
Thus the $q_1$-scattering between two Krames degenerate states is forbidden. Here $\mathcal{T}$ is the time-reversal operator with $\mathcal{T}^2=-1$, and $\hat{U}$ is a TR-independent operator representing the non-magnetic scattering potential. Therefore, the LDOS oscillation vanishes to leading order. Taking the next order into account, it decays algebraically as $1/r^2$.
For a convex CEC shown in Fig.~\ref{impurity}(b), there exist only two kinds of characteristic wave vectors: $\vec{q}_1$ along $\bar{\Gamma}$-$\bar{K}$ and $\vec{q}_2$ along $\bar{\Gamma}$-$\bar{M}$ direction. Both of them connect a pair of TRS states whose scattering is prohibited, thus the LDOS oscillation decays as $1/r^2$. For a hexagon-like CEC shown in Fig.~\ref{impurity}(c), there exist three kinds of characteristic wave vectors: $\vec{q}_1$, $\vec{q}_3$ along $\bar{\Gamma}$-$\bar{K}$ and $\vec{q}_2$ along $\bar{\Gamma}$-$\bar{M}$. Obviously, the $\vec{q}_1$ and $\vec{q}_2$ scattering are forbidden for they connect pairs of TRS states. The
$\vec{q}_3$ connects a pair of states at EP, which dominates the
QPI, and it decays as $1/r$. With the concave hexagram Fermi surface as shown in
Fig.~\ref{impurity}(d), there exist only two kinds of characteristic wave
vectors: $\vec{q}_1$ along $\bar{\Gamma}$-$\bar{K}$ and $\vec{q}_2$
along $\bar{\Gamma}$-$\bar{M}$ direction. $\vec{q}_1$ is forbidden, and $\vec{q}_2$ dominates the
QPI with decay index $1/r$. The characteristic wave vector and decay power of the QPI pattern is summarized in Table 2.

\begin{table}[htbp]
\begin{center}
\renewcommand{\arraystretch}{1.6}
\begin{tabular*}{8cm}
{@{\extracolsep{\fill}}ccc}
\hline
\hline
\ & wave vector & decay power
\\ \hline
$0<E<0.4\varepsilon^*$  & - & $1/r^2$ \\
$0.4\varepsilon^*<E<0.69\varepsilon^*$  & $\bar{\Gamma}$-$\bar{K}$, $\bar{\Gamma}$-$\bar{M}$ & $1/r^2$ \\
$0.69\varepsilon^*<E<1.45\varepsilon^*$  & $\bar{\Gamma}$-$\bar{K}$ & $1/r$ \\
$1.45\varepsilon^*<E<\varepsilon_{\mathrm{max}}$  & $\bar{\Gamma}$-$\bar{M}$ & $1/r$ \\
\hline
\hline
\end{tabular*}\label{Tab1}
\end{center}\caption{Characteristic wave vector and power laws of QPI pattern for a non-magnetic impurity. $\varepsilon_{\mathrm{max}}$ is the energy upper bound for the surface Dirac electron.}
\end{table}

The scattering theory of the EPs can better explain the experiments. In the scattering configuration for Bi$_2$Te$_3$ as shown in Fig.~\ref{impurity}(d), by numerical calculations we obtain the allowed scattering vector $q_2$, which varies linearly with the energy as $q_2=1.5\bar{k}$ (rather than $\sqrt{3}\bar{k}$~\cite{zhangtong2009prl}), where $\bar{k}$ is the length of $\bar{\Gamma}$-$\mathbf{k}_7$. Together
with the STM data, we deduce the
Dirac velocity along $\bar{\Gamma}$-$\bar{K}$ should be
$v=4.15\times10^5$~m/s (instead of $4.8\times10^5$~m/~\cite{zhangtong2009prl}), in agreement with ARPES result $v=4.05\times10^5$~m/s.~\cite{chen2009science}.
In Bi$_2$Se$_3$, the CEC is circle-like up to $0.22$~eV and
the characteristic scattering wave vector is always along the diameter of the circle.
Thus, we expect the Fourier transformation of LDOS on Bi$_2$Se$_3$ surface is ring-like.

\subsection{Experiments}
\label{nonmagneticexperiment}

Yazdani's group reported
their STM study of disorder scattering of SSs on Bi$_{1-x}$Sb$_x$ alloy.~\cite{roushan2009nature} They used the energy-resolved FT-STS to draw the interference wave vectors in $q$-space. By comparing with spin-resolved ARPES data,
a comprehensive analysis of QPI patterns indicated there were \emph{eight} allowed scattering wave vectors, excluding those connecting states of opposite spins. With the spin-selective scattering they explicitly proved the TRS of topological SSs in this material. More recently, STM experiments further demonstrated that the topological SSs could penetrate barriers while maintaining their extended
nature in Sb.~\cite{seo2010nature}

Xue's group reported the first direct STM imaging of standing waves on the Bi$_2$Te$_3$ (111) surface.~\cite{zhangtong2009prl}
The interference fringes are caused by scattering off non-magnetic Ag atoms of the
topological states. The Dirac dispersion relation of $E(k)$ of SSs is confirmed by the voltage dependent QPI patterns.
The experiment of Xue's group indicates the backscattering of SSs is completely suppressed, which is a direct proof of the topological nature of the SSs in Bi$_2$Te$_3$. It worths mentioning that the allowed interference wave vectors in Bi$_2$Te$_3$ come from the \emph{intravalley} scattering, i.e. the SSs are within the single Dirac cone; while in Bi$_{1-x}$Sb$_x$ alloy the scattering wave vectors are related to the \emph{intervalley} scattering between different Dirac cones.

Later, Yazdani's group reported the STM study of the QPI patterns on the surface of Ca-doped Bi$_2$Te$_3$ at several bias voltages.~\cite{beidenkopf2011np} The standing wave patterns in $q$-space exhibit \emph{six} strong peaks along the $\bar{\Gamma}$-$\bar{M}$ direction for SSs with high energies, and circular patterns with lower ones. This can be well understood on the basis of the Fermi surface shape and its associated spin texture. The warped Fermi surface in Bi$_2$Te$_3$ dominates scattering of SSs with high energies relative to the Dirac point.
In the low-energy region where the dispersion of SSs is conic and the scattering results in a simple circular
pattern in the Fourier-transformed QPI. The low energy
scattering pattern is also consistent with the helical spin
texture of the topological SSs that still allows all scattering
processes except for the direct backscattering.

\section{Classical Magnetic impurity}
\label{magnetic}

\subsection{Theory}
\label{magnetictheory}

The TRS in TIs is broken by doped magnetic impurities, so that in principle the backscattering process will occur, which is prohibited in TI with non-magnetic impurities only. Here we focus on the effects
of a classical magnetic impurity, which implies the Kondo physics can be ignored.

Suppose the scattering potential of an magnetic impurity is
\begin{equation}
\mathcal{H}_1 = V_{i} \sigma^i \delta(\mathbf{r})=J^{\mathrm{eff}}_i\langle S_i\rangle \sigma^i \delta(\mathbf{r}),
\end{equation}
where $\sigma^i$ $(i=x,y,z)$ are the Pauli matrices denoting the electron spin, $\langle S_i\rangle$ is the expectation value of the local magnetic moment, and $J^{\mathrm{eff}}_i$ denotes the effective exchange coupling between electron spin and local magnetic moment. As indicated by Eq.~(\ref{TRI}), with TRS, the backscattering is absent between a pair of TR-states of
$|\mathbf{k},\uparrow\rangle$ and $\left|-\mathbf{k},\downarrow\right\rangle$.
It is thus straightforward to prove the TRS-breaking potential such as the potential of magnetic impurities can remove such a constraint on backscattering, because
$$\mathcal{T}\sigma_i\mathcal{T}^{-1}=-\sigma_i,$$ and
\begin{eqnarray}\label{TRS}
\left\langle-\mathbf{k},\downarrow\right|\hat{U}\left|\mathbf{k},\uparrow\right\rangle \neq 0.
\end{eqnarray}
However, such backscattering between TR pairs can \emph{hardly} be observed in the LDOS spectra.~\cite{zhou2010prb} Instead of having $\delta\rho_{\uparrow}(\mathbf{q},\omega)=\delta\rho_{\downarrow}(\mathbf{q},\omega)$
in the non-magnetic impurity case, here we have $\delta\rho_{\uparrow}(\mathbf{q},\omega)\approx-\delta\rho_{\downarrow}(\mathbf{q},\omega)$.
Suppose the magnetic moment of the impurity is along the $z$ direction, then the spin-up electrons and spin-down
electrons experience opposite scattering potentials. Thus to the lowest order of perturbation theory, the scattering
amplitudes of the spin-up and spin-down electrons have opposite sign, so that the total interference pattern
of the charge density vanishes almost everywhere.
\begin{eqnarray}
\delta\rho^1(\mathbf{q},\omega)&\approx& \int\frac{d^2k}{4\pi}\mathrm{Tr}\left[G_0(\mathbf{k},\omega)\mathcal{H}_1G_0(\mathbf{k+q},\omega)\right]
\nonumber
\\
&=&\sum_{i}V_i\int\frac{d^2k}{4\pi}\mathrm{Tr}\left[G_0(\mathbf{k},\omega)\sigma_iG_0(\mathbf{k+q},\omega)\right]
\nonumber
\\
&=&0.
\end{eqnarray}
This argument no longer holds if higher orders of perturbation are considered. The higher order perturbations to the scattering amplitude would lead to similar results as in non-magnetic impurities.

\subsection{Experiments}
\label{magneticexperiment}
Recently, Madhavan's group reported the first TR violating scattering vectors in magnetically doped TI of Bi$_2$Fe$_{2-x}$Te$_3$ by using FT-STM dI/dV maps.~\cite{okada2011prl} Similar to previous studies,~\cite{zhangtong2009prl} a sixfold symmetric pattern with intensity centered along the $\bar{\Gamma}$-$\bar{M}$ directions are observed above $150$~meV. Remarkably, at low energies starting around $60$~meV, a new set of scattering vectors centered along $\bar{\Gamma}$-$\bar{K}$ directions have appears, not reported in prior studies. By comparing with ARPES results, they claimed that the new scattering channels arose from the forbidden backscattering. However, at the energy $140$~meV where scattering vector along $\bar{\Gamma}$-$\bar{K}$ emerges for the CEC similar to the hexagon in Fig.~\ref{impurity}(c). The scattering vector by a non-magnetic or magnetic impurity should be along the $\bar{\Gamma}$-$\bar{K}$ direction solely due to the shape of Fermi surface. Therefore, whether these new scattering channels arise from the forbidden backscattering or not is still \emph{unsettled}.

Later, Yazdani's group reported the standing wave patterns on Mn-doped Bi$_2$Te$_3$ and Bi$_2$Se$_3$.~\cite{beidenkopf2011np} The detailed Fourier
analysis of the QPI patterns in $q$-space shows that the short wavelength scattering in different samples have similar results, which is independent of whether the scattering dopant is magnetic or not, or even if magnetic order is established. The FT-QPI pattern consists of \emph{six} strong peaks along the $\bar{\Gamma}$-$\bar{M}$ directions at high energies and circular patterns at lower ones. These results strongly support that in the presence of weak magnetic impurities the backscattering between time-reversal pairs can hardly be seen in the LDOS.

Kimura's group reported the QPI pattern induced by cobalt adatoms on the Bi$_2$Se$_3$ surface by STM.~\cite{ye2012prb}
They find that Co atoms are selectively adsorbed on top of Se sites and act as strong scatter centers
on the surface, generating anisotropic standing waves along $\bar{\Gamma}$-$\bar{M}$ directions. However, the long-range ferromagnetic order is found to be absent, and the Dirac cone of SSs remains gapless. The anisotropy of the standing waves at high bias voltage is ascribed to the heavily warped
CEC of unoccupied states. At low voltage the QPI patterns due to the occupied states near Dirac cone vanishes, which suggests that the time-reversal protection of the topological SSs persists even in the presence of Co impurities.

\section{Other defects}
\label{otherimpurity}

Yandani's group studied the influence of the bulk-originated disorder
potential on the short-wavelength interference patterns caused by the surface impurities in Bi$_{1.95}$Mn$_{0.05}$Te$_3$.~\cite{beidenkopf2011np} They find the wave vector shift of the QPI pattern in the real-space LDOS with high and low voltages, which suggest the surface Dirac electrons alter their wavelength to adjust to the underlying bulk disorder potential, and are not immune to such perturbations.
Whereas such fluctuations in momentum are a relatively weak perturbation to the Dirac electrons with high energies; near the Dirac point they are comparable to the average value of the momentum. The lack of well-defined momentum near the
Dirac point due to the fluctuations reported is also likely to play a role in the apparent suppression of the ARPES measured signals in magnetically-doped TIs advocated recently.~\cite{chen2010science,wray2011np}

Madhavan's group used the STM and STS to investigate the effects of one-dimensional buckling on the electronic properties of Bi$_2$Te$_3$.~\cite{okada2012nm} By tracking spatial variations of the interference patterns generated by the Dirac electrons, they show that buckling imposes a periodic potential, which locally modulates the SS dispersion. Such potential induces the scattering along $\bar{\Gamma}$-$\bar{M}$ direction, which support the absence of backscattering and well-nesting of the snowflake-like CEC.

\section{Conclusion}
\label{conclusion}

The theoretical and experimental investigations
indicate that the LDOS oscillation on the surface of TIs is
generally determined by the scattering between surface states around the
extremal points on Fermi surface, either by step edges or by
non-magnetic impurities. The forbidden backscattering and suppressed standing wave oscillation clearly demonstrate the 2D Dirac nature of topological surface states. The robustness of the topological surface states may have potential application in quantum computing or spintronics in the future.

\acknowledgments
This work is supported by the Program of Basic Research Development of China (Grant No. 2011CB921901) and
National Natural Science Foundation of China (Grant No. 11074143).

\end{document}